
\input phyzzx
%
\Pubnum={UVA-HET-92-07\cr
cond-mat/9208016}
\date={August, 1992}
\titlepage
\title{Invariants of the Haldane-Shastry $SU(N)$ Chain}
\bigskip
\author {Michael Fowler\footnote\dag
{mf1i@landau1.phys.virginia.edu}
and
Joseph~A.~Minahan\footnote\ddag
{minahan@gomez.phys.virginia.edu}}
\address{Department of Physics,  Jesse Beams Laboratory,\break
University of Virginia, Charlottesville, VA 22901 USA}
\bigskip
\abstract{
Using a formalism developed by Polychronakos,
we explicitly construct a set of invariants of the motion for the
Haldane-Shastry $SU(N)$ chain.
}
\submit{Physical Review Letters, {\rm PACS numbers: 75.10.Jm, 05.30.-d,
03.65.Ca}}
\endpage

\def\PRA{{\it Phys. Rev. A\ }}
\def\PRB{{\it Phys. Rev. B\ }}
\def\PRL{{\it Phys. Rev. Lett.\ }}

\REF\HalShas{F.~D.~M.~Haldane, \PRL {\bf 60}, 635 (1988); B.~S.~Shastry,
\PRL {\bf 60}, 639 (1988).}
\REF\Ino{V.~I.~Inozemtsev, {\it Jour. Stat. Phys.}, {\bf 59}, 1143 (1989).}
\REF\Spinon{F.~D.~M.~Haldane, \PRL {\bf 66}, 1529 (1991).}
\REF\KA{H.~Kiwata and Y.~Akutsu, {\it Jour. Phys. Soc. Jap.},
{\bf 61}, 1441, (1992).}
\REF\Kaw{N.~Kawakami, \PRB {\bf 46}, 1005 (1992); \PRB {\bf46}, 3191 (1992).}
\REF\Shas{B.~S.~Shastry, \PRL {\bf 69}, 164 (1992).}
\REF\HaHal{Z.~N.~C.~Ha and F.~D.~M.~Haldane, Princeton preprint, 1992.  To
appear in Phys. Rev. B.}
\REF\Halpriv{F.~D.~M.~Haldane, private communication.}
\REF\Suth{B.~Sutherland, \PRA {\bf 4}, 2019 (1971); \PRA {\bf5}, 1372 (1972);
\PRL {\bf34}, 1083 (1975).}
\REF\app{A.~P.~Polychronakos, \PRL {\bf69}, 703 (1992).}
\REF\Calog{F.~Calogero, {\it Jour. of Math. Phys.}
{\bf10}, 2191 and 2197 (1969); {\bf 12}, 419 (1971).}
\REF\Moser{J.~Moser, {\it Adv. Math.} {\bf 16}, 1 (1975); F.~Calogero, {\it
Lett. Nuovo Cim.} {\bf13}, 411 (1975); F.~Calogero and C. Marchioro, {\it
Lett. Nuovo Cim.} {\bf13}, 383 (1975).}
\REF\jamapp{J.~A.~Minahan and A.~P.~Polychronakos, Virginia and Columbia
preprint UVA-HET-92-04, CU-TP-566, submitted to Phys. Rev. Lett.}

\def\half{{1\over2}}
\def\al{\alpha}

\def\Lam{\Lambda}

\def\Svec{\vec S}
\def\Lvec{\vec \Lambda}

\def\Mij{M_{ij}}
\def\zij{z_{ij}}
\def\zik{z_{ik}}
\def\zki{z_{ki}}
\def\zjk{z_{jk}}
\def\zkj{z_{kj}}
\def\zil{z_{il}}
\def\zjl{z_{jl}}
\def\zkl{z_{kl}}
\def\zlk{z_{lk}}
\def\Mkm{M_{km}}
\def\Mjk{M_{jk}}
\def\Mjl{M_{jl}}
\def\Mik{M_{ik}}
\def\Mil{M_{il}}
\def\Pij{P_{ij}}
\def\pii{\pi_i}
\def\pij{\pi_j}
\def\sigij{\sigma_{ij}}
\def\sigjk{\sigma_{jk}}
\def\sigkm{\sigma_{km}}
\def\sigv{\vec \sigma}

\def\sigp{\sigma^+}
\def\sigm{\sigma^-}
\def\sumij{\sum_{i\ne j}}
\def\Ht{\widetilde H}

\def\It{\tilde I}

There have been several recent papers on the Haldane-Shastry model for spin
chains and its $SU(N)$ generalization [\HalShas-\HaHal].
This model is described by the hamiltonian
$$H=\sum_{i<j}{1\over\sin^2{\pi\over L}(x_i-x_j)}\Pij,\eqn\scham$$
where $x_i$ are the positions of the spins, equally spaced around a ring,
and $\Pij$ is the operator that
exchanges the spins at sites $i$ and $j$.  Haldane and Shastry found the
wave functions for the antiferromagnetic ground state [\HalShas],
showing it to be
identical in form to the Sutherland ground state wave function
for particles on a line with the inverse square potential [\Suth].
These authors also found all possible energy levels for the system.

It would thus seem that the Haldane-Shastry model and its generalizations
are integrable systems.  If the model is integrable, there must
exist a set of operators that commute among themselves and with the
Hamiltonian.
Inozemtsev found the first such nontrivial operator, one
involving the exchange of three spins [\Ino].
Haldane later found two others, a four-spin-exchange operator that commutes
with
both the Hamiltonian and with Inozemtsev's operator, and a more basic
two-spin-exchange vector operator
he refers to as the rapidity [\Halpriv].

In this paper we explicitly show that the Haldane-Shastry model is integrable
by constructing a complete set of operators that commute among themselves
and with the
Hamiltonian.  These operators are very similar in structure to those
used by Polychronakos [\app] in his exchange operator approach to the
Sutherland and Calogero models [\Suth,\Calog,\Moser], and by Polychronakos
and one of the authors in generalizations of these models [\jamapp].

The key to our approach is that we consider the system as a set of $N$ bosons
with internal degrees of freedom which sit on the $N$ sites of the lattice,
only allowing states with one particle per site, as in the infinite-$U$
Hubbard model.  The exchange terms making up the Hamiltonian provide both the
kinetic energy, from hopping exchange of particles with different internal
quantum states, and the potential energy.  The new feature revealed by our
approach is that there are very simple single {\it particle} (as opposed to
single {\it site}) operators that commute with the Hamiltonian, analogous to
those used by Polychronakos for the continuum case [\app].
This makes it possible to construct a series of extensive conserved quantities.

To begin, let us assume that we have $N$ bosonic particles sitting at different
points on the circle.  Let us further assume that the system propagates only
by the particles exchanging their positions.  Therefore, if the system starts
with $N$ particles on $N$ different sites, the system will evolve with one
particle on each of these same $N$ sites.
If the bosons had no other quantum numbers besides their positions, then
this would be a trivial system.  However, if the bosons have internal
degrees of freedom then we will find a system with nontrivial dynamics.

Inspired by the work of Polychronakos [\app], we define an operator $\pii$,
$$\pii=\sum_{j\ne i}{z_j\over\zij}\Mij\eqn\pieq$$
where $z_i=\exp(2\pi i x_i/L)$, $x_i$ are the particle positions,
$\zij=z_i-z_j$ and $\Mij$ is the operator that
exchanges the positions of particles $i$ and $j$.  $\Mij$ is an hermitian
operator that satisfies the relations
$$\eqalign{&\Mij z_i=z_j\Mij,\qquad\qquad
\Mij z_k=z_k\Mij\qquad {\rm if}\ i\ne k\ne j,\cr
&M_{jik}\equiv\Mij\Mik=\Mjk\Mij=\Mik\Mjk.}\eqn\Mrel$$
Using these relations it is straightforward to show that
$$[\pii,\pij]=\Mij\pii-\pii\Mij,\eqn\picomm$$
and therefore
$$[\pii^n,\pij]=\Mij\pii^n-\pii^n\Mij.\eqn\pincomm$$
The operator $\pii$ is very similar to an operator considered by Polychronakos,
the only difference being that our operator does not contain an
explicit kinetic term.

Next consider the operator $\It_n$,
$$\It_n=\sum_i \pii^n.\eqn\Indef$$
Computing the commutator of $\It_n$ with $\It_m$ we find,
$$\eqalign{[\It_n,\It_m]&=\sum_{i,j}[\pi_i^n,\pi_j^m]=\sum_{i,j}\sum_{\al=0}^{m-1}
\pi_j^\al[\pi_i^n,\pi_j]\pi_j^{m-\al-1}\cr
&=-\sum_{i,j}\Biggl(\sum_{\al=0}^{m-1}-\sum_{\al=n}^{m+n-1}\Biggr)
\pi_j^\al\Mij \pi_j^{m+n-\al-1}.}\eqn\Incomm$$
Explicitly antisymmetrizing in $m$ and $n$ then gives
$$[\It_n,\It_m]=-\sum_{i,j}\Biggl(\sum_{\al=0}^{m-1}-\sum_{\al=n}^{m+n-1}
-\sum_{\al=0}^{n-1}+\sum_{\al=m}^{m+n-1}\Biggr)h_j^\al\Mij h_j^{m+n-\al-1}=0.
\eqn\Incommz$$
Note that the commutation of these operators does not need the spacing
between the sites to be equidistant.

We next relate the $\It_n$ operators to corresponding operators in the
Haldane-Shastry model.  Operators in the Haldane-Shastry model involve
the exchange of spins at particular sites on the lattice.  Our operators
involve the exchange of positions of particles that live on each site.
But we can invoke the fact that the particles are identical to relate
the two sets of operators [\jamapp].
Let us define an operator $\sigij$ that exchanges
the spins of two particles, but not their positions.  If the particles
are identical, then the product $\sigij\Mij$ acting on a symmetric
wave-function
is unity.  Moreover, since $\sigij$ acts on spins and $\Mij$ acts on
the positions, the two operators commute with each other.  Hence, if we
have a chain of $M$ operators acting on a symmetric state, we can substitute
for it a chain of spin exchange operators.  For example, we can make the
substitution on the following product of operators:
$$\Mij\Mjk\Mkm|\psi\rangle=\Mij\Mjk\sigkm|\psi\rangle=\sigkm\Mij\sigjk|\psi
\rangle=\sigkm\sigjk\sigij|\psi\rangle.\eqn\opprod$$
This product of $\Mij$ operators is the cyclic exchange operator for four
particles.  The corresponding spin product is the cyclic exchange of spins
in the opposite direction.

Finally, to relate these spin operators of
{\it particles} to spin operators at {\it sites}, we note that
there is always one
particle at every site.  Hence any operator that contains symmetric
sums over all
particles can be substituted with an operator that sums over all sites,
and where the spin exchange operator for {\it particles} $\sigij$, is replaced
by the exchange operator for {\it sites} $\Pij$.  In particular, the operators
$\It_n$ are now replaced with new operators $I_n$.
The commutation relations of the $I_n$ operators will be the same as the
commutation relations for the original operators.

To complete the proof of integrability, we consider the operator $\Ht$,
$$\eqalign{\Ht=&\sum_j\Ht_j\cr
\Ht_j=&\sum_{k\ne j}{1\over\sin^2{\pi\over L}(x_j-x_k)}\Mjk\cr
=&-4\sum_{k\ne j}{z_kz_j\over(\zkj)^2}\Mjk,}\eqn\pxham$$
We now show that all $\pi_i$ commute with $\Ht$, if the sites are
equally spaced.
Consider first the commutator of $\pii$ with $\Ht_j$ if $i\ne j$.
Using the relations in \Mrel, we find that
$$\eqalign{[\pii,\Ht_j]&=\sum_{k\ne i\atop l\ne j}\biggl[{z_k\over\zik}\Mik,
{-4z_jz_l\over (\zjl)^2}\Mjl\biggr]\cr
&=-4\sum_{k\ne i,j}\biggl\{
\biggl[{z_k\over \zik}\Mik,{z_jz_i\over (\zij)^2}\Mij\biggr]
+\biggl[{z_j\over \zij}\Mik,{z_jz_i\over (\zij)^2}\Mij\biggr]\cr
&\qquad\qquad\qquad
+\biggl[{z_k\over \zik}\Mik,{z_jz_k\over (\zjk)^2}\Mjk\biggr]\biggr\}\cr
&=-4\sum_{k\ne i,j}\biggl(
{z_k\over\zik}{z_jz_k\over(\zjk)^2}
+{z_j\over\zij}{z_iz_k\over(\zik)^2}
-{z_j\over\zij}{z_jz_k\over(\zjk)^2}\biggr)M_{ijk}\cr
&\qquad-4\sum_{k\ne i,j}\biggl(
-{z_k\over\zik}{z_jz_k\over(\zjk)^2}
-{z_k\over\zjk}{z_iz_j\over(\zij)^2}
-{z_k\over\zik}{z_jz_i\over(\zij)^2}\biggr)M_{jik}\cr
&=-4\sum_{k\ne i,j}\biggl\{-{z_iz_jz_k\over\zjk(\zik)^2}M_{ijk}
-{z_jz_kz_j\over\zij(\zjk)^2}M_{jik}\biggr\}.}\eqn\piihtj$$
Next consider the commutator of $\pii$ with $\Ht_i$.  We find
$$\eqalign{[\pii,\Ht_i]&=-4\sum_{k\ne i\atop l\ne i}\biggl[{z_k\over\zik}\Mik,
{z_iz_l\over (\zil)^2}\Mil\biggr]\cr
&=-4\sum_{k\ne i}\biggl[{z_k\over \zik}\Mik,{z_iz_k\over (\zik)^2}\Mik\biggr]
-4\sum_{k,l\ne i\atop k\ne l}\biggl[{z_k\over \zik}\Mik,
{z_iz_l\over (\zil)^2}\Mil\biggr]\cr
&=-4\sum_{k\ne i} {z_i+z_k\over\zik}{z_iz_k\over(\zik)^2}
-4\sum_{k,l\ne i\atop k\ne l}\biggl({z_kz_kz_l\over \zik(\zkl)^2}M_{kil}
-{z_iz_lz_k\over \zlk(\zil)^2}M_{ikl}\biggr).}\eqn\piihti$$
Summing over $j$ in \piihtj\ and adding the expression in \piihti,
we are left with
$$[\pii,\Ht]
=-4\sum_{k\ne i} {z_i+z_k\over\zik}{z_iz_k\over(\zik)^2}.\eqn\piiht$$
In general, this expression is not zero.  However, if we assume that the
sites are equally spaced, then by translational invariance and the antisymmetry
of the  summand, the sum is zero.

Since all $\pii$ commute with $\Ht$ then clearly, all $\It_n$ must commute
with $\Ht$
as well.  We may now perform the substitution of $H$ for $\Ht$ in the
same way that $I_n$ is substituted for $\It_n$.    Hence all $I_n$ commute
with $H$ and therefore the system is integrable.

Having established that the $I_n$ form a commuting set of operators, we now
examine some of these operators explicitly.  The first such operator, $I_1$,
is found from $\It_1$, which is given by
$$\It_1=\sum_{i\ne j}{z_j\over z_{ij}}M_{ij}=-\half\sum_{i\ne j}{\zij\over\zij}
\Mij.\eqn\Iti$$
Thus $I_1$ satisfies
$$I_1=-\half\sum_{i\ne j}\Pij=-{N(N-4)\over4}-(\Svec\cdot\Svec),\eqn\Ii$$
where $\Svec$ is the total spin of the system.  This operator trivially
commutes
with the Hamiltonian.

A more interesting operator is $I_2$, where $\It_2$
is given by
$$\It_2=\sum_i\sum_{j\ne i\atop k\ne i}{z_j\over\zij}\Mij {z_k\over\zik}\Mik
=\sum_{i\ne j\ne k\ne i}{z_j\over\zij} {z_k\over\zjk}M_{ikj}+
\sum_{i\ne j}{z_i z_j\over(\zij)^2}.\eqn\Itii$$
The last term is just a constant.  Symmetrizing the sum
over $i$, $j$, and $k$, we find that $\It_2$ reduces to
$$\It_2=\half\sum_{i\ne j\ne k\ne i}{z_i+z_j\over\zij}M_{ijk}+{1\over6}
\sum_{i\ne j\ne k\ne i}M_{ijk}-{1\over12}(N^2-1).\eqn\ItII$$
Hence we have
$$I_2=-\half\sum_{i\ne j\ne k\ne i}{z_i+z_j\over\zij}P_{ijk}
+{1\over6}\sum_{i\ne j\ne k\ne i}P_{ijk}-{1\over12}(N^2-1).\eqn\Iii$$
The second term is a trivial exchange operator that commutes with the
Hamiltonian and the other $I_n$, therefore the first term must do
so as well.  To demonstrate the significance of this operator, let us
specialize to the case of $SU(2)$.  The antisymmetric piece of $P_{ijk}$
is given by $-i(\sigv_i\times\sigv_j)\cdot\sigv_k$.  Since the first term
in \Iii\ has no $z_k$ dependence, it can be reexpressed as
$${i\over2}\sum_{i\ne j}{z_i+z_j\over\zij}(\sigv_i\times\sigv_j)\cdot\Svec
=\Lvec\cdot\Svec.\eqn\III$$
Since the Hamiltonian is isotropic in the total spin, then each component of
$\Lvec$ must commute with $H$.  This operator is the rapidity operator
defined by Haldane [\Halpriv].  Let us act with this operator on
the one magnon state, described by the wave function
 $$|\psi\rangle=\sum_n\sigp_ne^{ikn}|0\rangle,\eqn\magnon$$
where $|0\rangle$ is the all spins down state.  Acting on this state with
$\Lam_z$ gives
$$\eqalign{
&\half\sumij{z_i+z_j\over\zij}{1\over2}(\sigm_i\sigp_j-\sigp_i\sigm_j)
\sum_n(z_n)^{k}\sigp_n|0\rangle\cr
&={1\over2}\sumij{z_i+z_j\over\zij}\left({z_j\over z_i}\right)^{k}
|\psi\rangle\cr
&={1\over2}N\sum_{j\ne0}{1+z_j\over1-z_j}(z_j)^{k}|\psi\rangle\cr
&=N(k-N/2)|\psi\rangle.}\eqn\magmo$$
Hence the $z$ component of $\Lvec$ acts like a momentum operator.

For higher $I_n$, one can show that the leading term is of the form
$$I_n\propto\sum_{i_1\ne....i_{n+1}}{z_{i_1}+z_{i_2}\over z_{i_1}-z_{i_2}}
{z_{i_2}+z_{i_3}\over z_{i_2}-z_{i_3}}......
{z_{i_{n-1}}+z_{i_n}\over z_{i_{n-1}}-z_{i_n}}P_{i_1...i_{n+1}}.\eqn\Ineq$$
This is basically a generalization of the rapidity operator and is
not quite of the Inozemtsev-Haldane form for invariants.
Since the leading term in $I_n$ contains an $n$-spin exchange term, it must
be independent of all $I_m$, $m<n\le N$, since $I_m$ won't have such a term.

Presumably, the Hamiltonian as well as the
Inozemtsev-Haldane invariants lurk within our operators, but they fit
in a nontrivial way.  For instance, after a particularly tedious calculation
one can show that $I_3$ is given by
$$\eqalign{I_3=&\Lvec_2\cdot\Svec-{N-3\over4}\Lvec\cdot\Svec
-\Lvec\cdot\Lvec+{3\over4}H\cr
&-{N^2+3N+20\over24}\sumij\Pij+{3N-14\over24}\sum_{i\ne j\ne
k}P_{ijk}-{1\over8}\sum_{i\ne j\ne k\ne l}P_{ijkl}+C,}\eqn\Iiii$$
where $\Lvec_2$ is
$$\Lvec_2=-\half
\sum_{i\ne j\ne k}{z_i+z_j\over\zij}{z_j+z_k\over\zjk}P_{ijk}\sigv_k.
\eqn\Lvii$$
The Hamiltonian explicitly appears in $I_3$, but one can also show that the $z$
component of $\Lvec_2$ acting on the one magnon state satisfies
$$\Lambda_{2z}|\psi\rangle=-{1\over4}\biggl(H-(N-1)\sumij\Pij
+{1\over3}(N^2-5)\biggr) |\psi\rangle.\eqn\magii$$
The Hamiltonian and $\Lvec\cdot\Lvec$ acting on the one magnon state lead
to terms quadratic in the momentum, hence all terms in $I_3$ are basically
equivalent to terms containing the Hamiltonian or the rapidity, at least when
acting on one magnon states.
Likewise $I_4$ will contain
Inozemtsev's operator [\Ino],
$$\sum_{i\ne j\ne k}{z_iz_jz_k\over\zij\zjk\zki}P_{ijk},\eqn\Inop$$
and other operators that lead to terms cubic in the momentum when acting
on a single magnon.

In conclusion, we have shown that the Haldane-Shastry $SU(N)$ chain is
integrable by explicitly constructing a set of independent invariants of the
motion.
For the discrete
case considered here, the Hamiltonian appears in the third level of invariants.
This contrasts to the Sutherland continuum model, where the Hamiltonian first
appears in the second level of invariants.
In general, $I_n$ acting on one magnon states will give $n-1$ powers of the
momentum.  Hence the $I_n$ are like
derivative operators, although there is one less derivative than in the
continuum case.

\ack{We thank F.D.M. Haldane for communicating his results prior to
publication.   M.F. was supported in part by NSF grant DMR-88-10541.
J.A.M. was supported in part by
D.O.E. grant DE-AS05-85ER-40518.}

\refout
\end